\documentclass[sigconf]{acmart}

\setcopyright{acmlicensed}
\copyrightyear{2018}
\acmYear{2018}
\acmDOI{XXXXXXX.XXXXXXX}
\acmConference[Conference acronym 'XX]{Make sure to enter the correct
  conference title from your rights confirmation email}{June 03--05,
  2018}{Woodstock, NY}
\acmISBN{978-1-4503-XXXX-X/2018/06}

\usepackage{booktabs}
\usepackage{amsmath}
\usepackage{float}
\usepackage{multirow}

\begin{document}

\title{UGR: Unified Generative Ranking for Heterogeneous E-commerce Recommendation}

\settopmatter{authorsperrow=4}

\author{Zonghe Shao}
\authornotemark[1]
\affiliation{
  \institution{Alibaba Group}
  \city{Hangzhou}
  \country{China}}
\email{zhshao@hust.edu.cn}

\author{Huabin Chen}
\authornote{Both authors contributed equally to this research.}
\affiliation{
  \institution{Alibaba Group}
  \city{Hangzhou}
  \country{China}}
\email{chenhuabin.chb@taobao.com}

\author{Sitong Lu}
\affiliation{
  \institution{Alibaba Group}
  \city{Hangzhou}
  \country{China}}
\email{dailan.lst@taobao.com}

\author{Fangmei Zhu}
\affiliation{
  \institution{Alibaba Group}
  \city{Hangzhou}
  \country{China}}
\email{fangmei.zfm@taobao.com}

\author{Yingcai Ma}
\affiliation{
  \institution{Alibaba Group}
  \city{Hangzhou}
  \country{China}}
\email{yingcai.myc@taobao.com}

\author{Ge Fan}
\affiliation{
  \institution{Alibaba Group}
  \city{Hangzhou}
  \country{China}}
\email{fange.fg@taobao.com}

\author{Jinguo Gao}
\affiliation{
  \institution{Alibaba Group}
  \city{Hangzhou}
  \country{China}}
\email{gaojinguo.gjg@alibaba-inc.com}

\author{Wei Shi}
\affiliation{
  \institution{Alibaba Group}
  \city{Hangzhou}
  \country{China}}
\email{shiwei.shiw@alibaba-inc.com}

\author{Yanfang Liu}
\affiliation{
  \institution{Alibaba Group}
  \city{Hangzhou}
  \country{China}}
\email{liuyanfang.lyf@alibaba-inc.com}

\author{Huiping Chu}
\authornote{Corresponding author.}
\affiliation{
  \institution{Alibaba Group}
  \city{Hangzhou}
  \country{China}}
\email{yueqi.chp@taobao.com}

\renewcommand{\shortauthors}{Shao et al.}

\begin{abstract}
List-wise ranking in e-commerce recommendation must balance user utility and commercial value across organic content, ads, and candidates from other businesses, while modeling candidate interactions and sequential dependencies.
Existing generative ranking methods typically learn from logged exposure sequences and operate on truncated candidate sets from a single business. Exposure-sequence learning may not fully exploit post-exposure feedback signals. Meanwhile, generative ranking over heterogeneous, multi-business candidates remains less explored, particularly when recommendation scores and advertising value must be jointly modeled under generalized second-price (GSP) auctions.
We propose \textbf{Unified Generative Ranking (UGR)}, which performs generative ranking over the full request-level multi-business heterogeneous candidate set, typically comprising around a thousand candidates in production, and directly generates multi-business ranking lists through candidate-aware autoregressive decoding. 
To go beyond exposure-sequence imitation, UGR introduces complementary feedback supervision through weighted supervised fine-tuning and permutation preference optimization, which respectively leverage observed post-exposure feedback and feedback-derived preferences among permutations.
UGR uses a training-time advertising debiasing tower to model historical ad exposure effects and incorporates current ad value during serving while preserving the existing allocation and GSP pricing rules.
Multi-value sequence retrieval produces candidate ranking lists, and the existing downstream evaluator selects the final list for exposure.
Offline experiments demonstrate consistent gains in ranking quality, while online A/B tests on Taobao’s Guess You Like recommendation service yield improvements of \textbf{1.76\%} in Core IPV, \textbf{3.29\%} in Transaction Volume, and \textbf{2.10\%} in GMV over the production baseline.

\end{abstract}



\maketitle

\section{Introduction}
\label{sec:introduction}

E-commerce recommendation selects a limited number of items for display from a heterogeneous candidate set containing organic content, advertisements (ads), and candidates from other businesses~\cite{carrion2023blending}.
These candidates differ in their features, feedback signals, and value objectives.
Organic content is primarily ranked according to user preferences, while advertisements additionally involve the advertising value associated with bids~\cite{gemrec}. 
An item’s value also varies with its surrounding candidates and display position ~\cite{prm,seq2slate}.
Industrial ranking must therefore model candidate interactions while balancing user utility and commercial value~\cite{cmr}.

\begin{figure}[t]
    \centering
    \includegraphics[width=\linewidth]{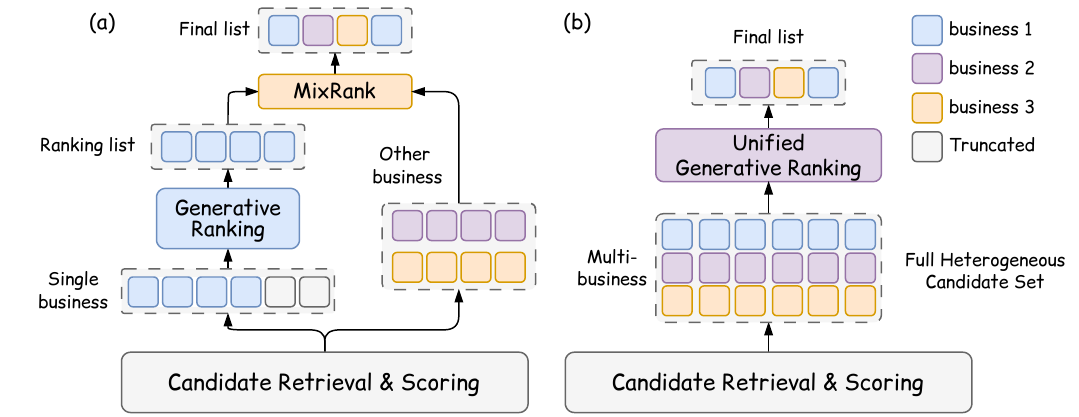}
    \caption{Generation workflows. 
        (a) Single-business generation followed by cross-business blending;
        (b) Unified generation over the full heterogeneous candidate set.
        }
    \label{fig:fig1}
\end{figure}

Prior studies progressed from modeling candidate interactions in discriminative item-level scoring~\cite{dlcm, prm, setrank} to constructing lists through sequential candidate selection~\cite{seq2slate, grn}. 
Generative ranking~\cite{huang2025towards,guo2025twin} further shifts ranking from candidate scoring toward list generation, strengthening the modeling of list context and dependencies among candidates through feedback-based learning~\cite{gref, goalrank, degre} and generation mechanisms~\cite{nar4rec, dcdr, unirank}. 
However, these methods typically operate on truncated, single-business candidate sets~\cite{gref, dualrerank}. 
In industrial settings requiring cross-business blending, the generated ranking list is subsequently combined with items from other businesses through mixrank to produce the final list. 
This blending changes item positions and list context, creating a discrepancy between the generated ranking list and the final list, as illustrated in Fig.~\ref{fig:fig1}(a). 
This discrepancy motivates a unified generation over the full request-level heterogeneous candidate set, which jointly determines the selected items, their ranking, and the composition of the final list, without subsequent cross-business blending (Fig.~\ref{fig:fig1}(b)).

However, deploying this approach in industrial systems presents three key challenges.
First, \textbf{efficiently modeling interactions over the full heterogeneous candidate set}.
In production, each request typically contains on the order of a thousand candidates spanning multiple businesses. 
These candidates compete for the same display positions, while their heterogeneous features and value objectives complicate list-context modeling.
Efficient contextualization is therefore required under online serving latency constraints.
Second, \textbf{learning user preferences beyond exposure-sequence imitation.}
Historical exposure sequences are shaped by existing ranking policies and do not fully reflect user preferences.
Post-exposure feedback reveals users' responses to exposed items, but effectively incorporating such feedback into generative ranking remains challenging.
Third, \textbf{enabling generative ranking under existing auction and pricing rules.}
The production system additively combines recommendation score and ad value, using the resulting score for both candidate competition and slot-wise pricing under the generalized second-price (GSP)~\cite{edelman2007gsp}  auction.
Generative ranking must incorporate current ad value while following this scoring convention without changing the existing GSP pricing rule.

To address these issues, we propose \textbf{UGR}, a \textbf{Unified Generative Ranking} framework that directly generates multi-business ranking lists from the full request-level heterogeneous candidate set received from the upstream candidate retrieval and scoring stages.
UGR encodes user attributes, request context, and historical behaviors into a request context token, and maps candidate features and business-type attributes into a common representation space.
Low-rank candidate attention contextualizes all request-level candidates, after which a candidate-aware autoregressive decoder produces prefix-conditioned candidate scores for list generation.
Offline training jointly optimizes next-token prediction (NTP), weighted supervised fine-tuning (WSFT), and permutation preference optimization (PermPO).
NTP learns logged exposure sequences, WSFT assigns greater training weight to exposed items
with higher observed feedback value, and PermPO uses feedback-based DCG to construct preference pairs over permutations of the same exposed items and provide relative ranking supervision.
UGR separates historical ad exposure modeling from current ad value integration. 
A training-time advertising debiasing tower models historical ad exposure effects and is removed during serving. 
Current ad value is then additively combined with recommendation scores in accordance with the existing allocation and pricing rules. 
Multi-value sequence retrieval uses the integrated scores to produce candidate ranking lists, from which the existing downstream evaluator selects the final list for exposure.
Offline experiments and online A/B tests on Taobao's Guess You Like recommendation service demonstrate UGR's effectiveness, with gains of \textbf{1.76\%} in Core IPV, \textbf{3.29\%} in Transaction Volume, and \textbf{2.10\%} in GMV over the production baseline.

Our main contributions are as follows:
\begin{itemize}
  \item We propose UGR,  a unified generative ranking framework that performs prefix-conditioned autoregressive generation over the full request-level heterogeneous candidate set, producing multi-business ranking lists without subsequent cross-business blending.

  \item We introduce complementary feedback supervision beyond exposure-sequence imitation by combining WSFT with PermPO. WSFT weights exposed items by observed feedback, while PermPO derives relative ranking supervision from feedback-based DCG over permutations of the same items.

  \item We use a training-time advertising debiasing tower to model historical ad exposure effects and additively combine current ad value with recommendation scores for multi-value sequence retrieval during serving, while preserving the existing allocation and GSP pricing rules.

\end{itemize}

\section{Related Work}
\label{sec:related_work}
\subsection{List-wise Generative Ranking}
\label{sec:related_generative_ranking}
List-wise ranking models candidate interactions and sequential dependencies during list construction. 
Context-aware approaches~\cite{dlcm,prm,setrank} encode the candidate list to produce contextualized item scores, while direct list generation approaches~\cite{seq2slate,grn} construct lists through successive candidate selection. 
Recent generative ranking studies improve generation efficiency through parallel decoding~\cite{nar4rec}, multi-token prediction~\cite{gref}, and knowledge distillation~\cite{dualrerank}, or enhance list construction through candidate-grounded generation~\cite{unirank}, globally consistent item representations~\cite{glorank, feng2026psg}, and diffusion-based permutation exploration~\cite{dcdr}.
However, existing generative methods typically operate on truncated, single-business candidate sets.
Subsequent cross-business blending can change the positions and context of the generated items, creating a discrepancy between the generated ranking list and the final list. 
UGR extends generative ranking to the full heterogeneous candidate set, thereby generating heterogeneous, multi-business ranking lists without subsequent cross-business blending.

\subsection{Reward and Preference Alignment}
\label{sec:related_preference_alignment}

Learning list dependencies from exposure logs provides a basis for generation, but exposure-sequence imitation alone cannot fully leverage post-exposure feedback.
Existing studies align generation with feedback or reward signals through relative list preferences~\cite{gref,degr}, estimated list utility~\cite{goalrank}, comparisons between neighboring lists~\cite{nlgr}, dense stepwise supervision~\cite{degre}, and value-aware learning objectives~\cite{gr4ad}.
Related studies further define list value using browsing behavior~\cite{swim} and cumulative click and conversion utility across e-commerce pages~\cite{nextgen}.
However, jointly incorporating item-level feedback supervision and list-level preference learning into multi-business generative ranking remains less explored.
UGR combines these complementary forms of supervision to learn list ranking beyond exposure-sequence imitation.

\subsection{Advertising Value Integration}
\label{sec:related_advertising_value}

Advertising value integration extends recommendation ranking beyond user utility to also consider advertising returns~\cite{li2026constraint}. 
Controllable multi-objective reranking~\cite{cmr} balances competing objectives during list construction, while generative auction approaches~\cite{genauction} optimize advertising allocation at the list level through generation and evaluation. 
Recent work incorporates ad value into generative recommendation through value-aware training~\cite{gr4ad} and bid-aware decoding for mixed organic and ad recommendations~\cite{gemrec}. 
These studies primarily focus on incorporating advertising value into ranking objectives and list generation. However, less attention has been paid to the effects of historical ad-serving policies reflected in recommendation scores learned from exposure logs. 
UGR separates historical ad exposure modeling during training from current ad value integration during serving, while following the existing allocation and GSP pricing rules.

\section{Problem Formulation}
\label{sec:problem_formulation}

\subsection{Generative Ranking}
\label{sec:generative_ranking_formulation}

Given a user and request context $x$, let $C=\{c_1,\ldots,c_M\}$ be the full request-level candidate set supplied by the upstream candidate retrieval and scoring stages. 
The set contains organic content, ads, and candidates from other businesses. 
Unified generative ranking constructs a ranking list $Y=(y_1,\ldots,y_L)$ directly from the candidate set $C$ for multi-business recommendation.
Let $y_{<t}=(y_1,\ldots,y_{t-1})$ be the generated prefix at step $t$, and let $C_t(y_{<t})$ denote the candidates not yet selected. 
The generative ranking model assigns a probability to each candidate in $C_t(y_{<t})$ conditioned on the request, the full candidate set, and the prefix:
\begin{equation}
p_{\theta}(Y\mid x,C)
=\prod_{t=1}^{L}
p_{\theta}(y_t\mid y_{<t},x,C),\quad y_t\in C_t(y_{<t})
\label{eq:autoregressive_ranking}
\end{equation}
where $\theta$ denotes the generator parameters.

\begin{figure}[t]
    \centering
    \includegraphics[width=\linewidth]{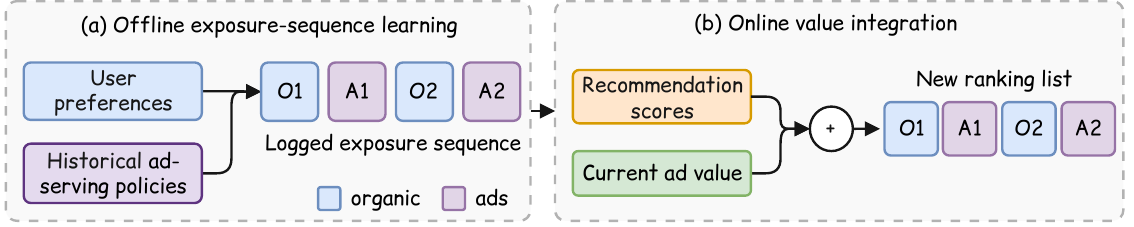}
    \caption{
    Historical advertising exposure effects. 
    (a) Logged exposure lists reflect both user preferences and the selection effects of historical ad-serving policies. 
    (b) Direct exposure fitting may carry historical ad exposure effects into recommendation scores, which are combined with current ad value during serving.
    }
    \label{fig:historical_ad_influence}
\end{figure}

\subsection{Advertising Considerations}
\label{sec:ad_allocation_pricing}
The production system uses the generalized second-price (GSP) auction~\cite{edelman2007gsp}, under which a winning ad pays the minimum bid required to retain its position in slot-wise candidate competition. 
With other competitive conditions fixed, increasing an ad’s bid should not worsen its allocation outcome.
Accordingly, the allocation and pricing rules require the ranking system to additively combine recommendation score and ad value, with the resulting score used for both candidate competition and slot-wise pricing.
Historical exposures, however, reflect both user preferences and the selection effects of the historical ad-serving policies~\cite{schnabel2016treatments}, including ad bids and competition.
As illustrated in Fig.~\ref{fig:historical_ad_influence}, directly fitting exposure lists can therefore embed historical ad exposure effects in recommendation scores, which are subsequently combined with current ad value when generating a new ranking list.
The ranking model therefore needs to prevent historical ad exposure effects from entering the recommendation scores used for online allocation.

The additive combination is formulated as follows.
Let $z_{t}(i)$ denote the generator output used as the recommendation score for candidate $i$ at position $t$. 
For an ad candidate, its ad value is $v_i = \mathrm{eCPM}_i = q_i b_i$, where $q_i$ is the predicted click-through rate and $b_i$ the advertiser's bid. 
The integrated score is $S_t(i) = z_{t}(i) + \alpha\, v_i$, where $\alpha > 0$ controls the contribution of ad value, and $v_i = 0$ for non-ad candidates.
The highest competing score is computed over all other legal candidates, including non-ad candidates.

\section{Methodology}
\label{sec:methodology}
\begin{figure*}[t]
    \centering
    \includegraphics[width=0.9\linewidth]{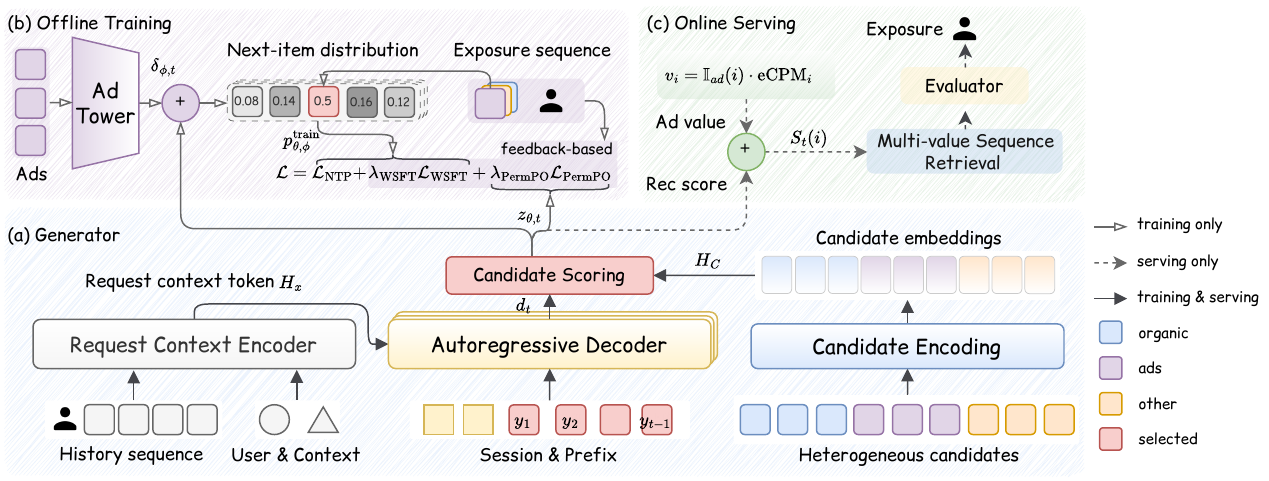}
    \caption{
    Overview of the UGR framework.
    (a) The generator performs request context encoding, heterogeneous candidate encoding, autoregressive decoding, and candidate scoring.
    (b) Offline training jointly optimizes NTP and WSFT using next-item distribution and PermPO using generator logits only. 
    (c) Online serving combines recommendation scores with current ad value for multi-value sequence retrieval, followed by the downstream evaluator. 
    }
    \label{fig:main}
\end{figure*}

Fig.~\ref{fig:main} summarizes UGR.
The generator autoregressively constructs lists over the full heterogeneous candidate set supplied by the upstream stage, while WSFT and PermPO provide feedback supervision beyond exposure imitation.
A training-time advertising debiasing tower models historical ad exposure effects and is removed during serving, where current ad value is integrated.

\subsection{Model Architecture}
\label{sec:model_architecture}

\subsubsection{Request Context Encoder}
\label{sec:encoder}
The request encoder embeds categorical features, transforms and projects numerical features, and represents historical behavior sequences across time scales as item tokens in recorded order. 
Masked self-attention~\cite{vaswani2017attention} aggregates the encoded user, request, and behavior features, and a learned token-axis projection compresses them into a request context token:
\begin{equation}
H_x=\mathrm{Encoder}_\theta(X_{\mathrm{user}},X_{\mathrm{context}},X_{\mathrm{history}})
\label{eq:request_encoding}
\end{equation}
where $X_{\mathrm{user}}$, $X_{\mathrm{context}}$, and $X_{\mathrm{history}}$ denote the user attributes, request context, and historical behavior inputs, respectively.

\subsubsection{Heterogeneous Candidate Encoding}
\label{sec:candidate_encoding}

UGR encodes candidates from different businesses through a common representation interface. 
Each candidate combines identity and content features, business-type attributes, and candidate-side prediction and ranking signals,
\begin{equation}
    e_i=\operatorname{Candidate\_Embed}_{\theta}(f_i,\tau_i)
    \label{eq:candidate_encoding}
\end{equation}
where $f_i$ contains candidate features and $\tau_i$ contains business-type attributes. 
This mapping retains business-specific information while placing heterogeneous candidates in a common embedding space for joint contextualization and scoring.

To model interactions across the full request-level set, UGR uses low-rank candidate attention~\cite{wang2020linformer}.
Let $E=[e_1,\ldots,e_M]^{\mathsf T}\in\mathbb R^{M\times D}$ stack the embeddings of the $M$ request-level candidates.
We apply RMSNorm~\cite{zhang2019rmsnorm} independently to each candidate embedding and compute queries and candidate-axis-compressed keys and values:
\begin{equation}
\begin{aligned}
    H&=\mathrm{RMSNorm}(E)\\
    Q=HW_Q,
    \bar K&=P_K(HW_K),
    \bar V=P_V(HW_V)\\
    A&=\mathrm{softmax}\!\left(
    Q\bar K^{\mathsf T}/\sqrt d
    \right)\bar V
\end{aligned}
\end{equation}
where $d$ is the attention dimension, and $P_K,P_V\in\mathbb R^{r\times M}$ are learned candidate-axis projections, with $r$ denoting the number of latent key–value components.
Residual connections and a feed-forward sublayer produce the contextualized representations:
\begin{equation}
    E'=E+A,\quad
    H_C=E'+\mathrm{FFN}
    \bigl(\mathrm{RMSNorm}(E')\bigr)
\end{equation}

The resulting $H_C=[h_1,\ldots,h_M]^{\mathsf T}$ contains the contextualized representations of the $M$ candidates, where each $h_i$ incorporates interactions with the full heterogeneous candidate set.

\subsubsection{Autoregressive Decoder}
\label{sec:decoder}

The decoder starts from the recent session tokens and appends the contextualized representation of each selected candidate. 
At position $t$, causal self-attention processes this sequence, while cross-attention uses $H_x$ as keys and values to incorporate user and request information.
Position-conditioned decoding produces the decoder state:
\begin{equation}
    d_t=\mathrm{Decoder}_\theta(X_{\mathrm{init}},[h_{y_1},\ldots,h_{y_{t-1}}],H_x)
\end{equation}
where $X_{\mathrm{init}}$ denotes the session context. 
A position-specific projection maps $d_t$ into a candidate-scoring vector $u_t$, and the generator logits are computed as scaled cosine similarities~\cite{zhang2019adacos}:
\begin{equation}
    z_{\theta,t}(i)=s_t\,\frac{u_t^{\mathsf T}h_i}{\|u_t\|_2\|h_i\|_2},\qquad u_t=\mathrm{Proj}_{\theta,t}(d_t)
\end{equation}
where $s_t$ is a learned logit scale.
The generator logits \(z_{\theta,t}(i)\) serve as recommendation scores and are normalized via softmax over the legal candidate set to obtain the next-item distribution:
\begin{equation}
    \label{eq:rec logits}
    p_\theta(i\mid y_{<t},x,C)=\frac{\exp z_{\theta,t}(i)}{\sum_{j\in C_t(y_{<t})}\exp z_{\theta,t}(j)},\qquad i\in C_t(y_{<t})
\end{equation}

The selected candidate representation is appended to the decoder input, and the legal set $C_t(y_{<t})$ is updated before the next step. Candidate interactions are encoded once in $H_C$, whereas their relevance to the evolving list is determined by $u_t$.

\subsubsection{Training-Time Advertising Debiasing Tower}
\label{sec:advertising_debiasing}
The generator in Fig.~\ref{fig:main}(a) represents ad and non-ad candidates using the same standard item features, excluding eCPM and other auction-derived features.
To model historical ad exposure effects separately, we introduce an advertising debiasing tower $g_{\phi}$ that takes ad item features and logged eCPM as input.
The tower produces an ad-specific logit adjustment at each decoding position:
\begin{equation}
\delta_{\phi,t}(i)=\mathbb I_{\mathrm{ad}}(i)\left[g_\phi(a_i;p_t)+\gamma_t\right], \quad a_i = [f^{(\text{ad})}_i, \text{eCPM}_i]
\label{eq:debiasing_tower}
\end{equation}
where $p_t$ is a learned position offset in the tower's hidden layer, and $\gamma_t$ is a position-specific scalar offset. 
The MLP weights are shared across positions, while the position offsets allow the adjustment to vary with the decoding position.
The indicator $\mathbb{I}_{\mathrm{ad}}(i)$ ensures that $\delta_{\phi,t}(i) = 0$ for non-ad candidates.
The tower is used only during training and is removed during serving.
During training, the tower adjustment is added to the generator logit to define the training next-item distribution over legal candidates:
\begin{equation}
p^{\mathrm{train}}_{\theta,\phi}(i\mid y_{<t},x,C)=\frac{\exp\left(z_{\theta,t}(i)+\delta_{\phi,t}(i)\right)}{\sum_{j\in C_t(y_{<t})}\exp\left(z_{\theta,t}(j)+\delta_{\phi,t}(j)\right)}
\label{eq:training_distribution}
\end{equation}

\subsection{Training Objectives}
\label{sec:training_objectives}

\begin{figure}[t]
    \centering
    \includegraphics[width=\linewidth]{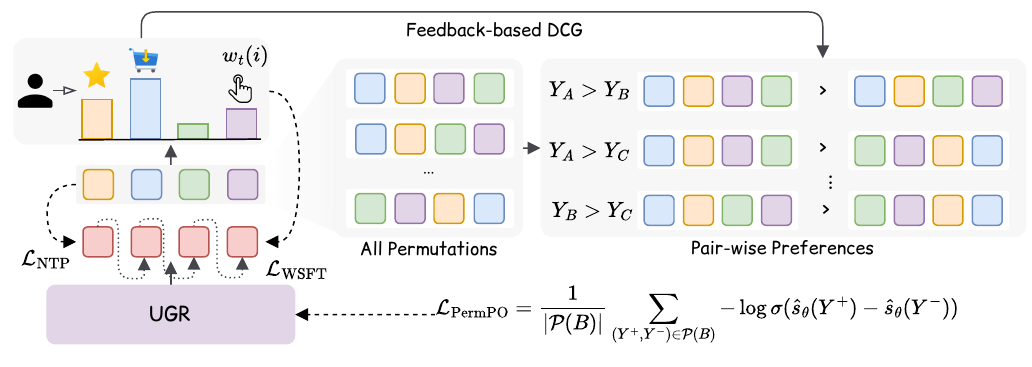}
    \caption{Complementary training objectives of UGR, including NTP, WSFT and PermPO.
    }
    \label{fig:train}
\end{figure}

As illustrated in Fig.~\ref{fig:train}, UGR combines three objectives. 
NTP provides baseline supervision from logged exposure sequences.
Beyond exposure imitation, WSFT and PermPO use observed feedback in complementary ways.
WSFT reweights the next-item log-likelihood term of each exposed item according to its observed feedback, whereas PermPO provides relative ranking supervision by comparing permutations of the same exposed items.

\subsubsection{Next Token Prediction}
\label{sec:ntp}

Given an observed exposure list $Y^{\mathrm{obs}}$, NTP predicts each exposed candidate conditioned on its observed prefix.
At each step, the probability of the observed next item is normalized over all legal candidates.
For the training data $\mathcal D$, the objective is
\begin{equation}
\mathcal{L}_{\mathrm{NTP}}
=-\mathbb{E}_{\mathcal{D}}\left[
\sum_{t=1}^{|Y^{\mathrm{obs}}|}
\log p^{\mathrm{train}}_{\theta,\phi}
\left(y_t^{\mathrm{obs}}\mid y_{<t}^{\mathrm{obs}},x,C\right)
\right]
\label{eq:ntp_loss}
\end{equation}

\subsubsection{Weighted Supervised Fine-Tuning}
\label{sec:wsft}
WSFT reweights the next-item log-likelihood of each exposed item  in NTP according to the observed feedback, while leaving the logged sequence and its prefixes unchanged.
Let $F_k(i)$ denote a nonnegative feedback signal associated with exposed item $i$, and define
\begin{equation}
    w(i)=\eta\sum_{k=1}^{K}\beta_k F_k(i),\qquad \eta>0,\quad \beta_k\geq0
    \label{eq:weight feedback}
\end{equation}

The feedback signals capture engagement and transaction outcomes.
Their definitions are provided in Appendix~\ref{app:training_configuration}.
With $w_t=w(y_t^{\mathrm{obs}})$, the objective is
\begin{equation}
\mathcal{L}_{\mathrm{WSFT}}
=-\mathbb{E}_{\mathcal{D}}\left[
\sum_{t=1}^{|Y^{\mathrm{obs}}|}
 w_t\log p^{\mathrm{train}}_{\theta,\phi}
\left(y_t^{\mathrm{obs}}\mid y_{<t}^{\mathrm{obs}},x,C\right)
\right]
\label{eq:wsft_loss}
\end{equation}

The NTP retains baseline supervision for all exposed items, including those with zero WSFT weight. 

\subsubsection{Permutation Preference Optimization}
\label{sec:permpo}
WSFT changes the contribution of observed next-item predictions but does not compare alternative rankings.
PermPO complements WSFT by deriving preferences between permutations of the same exposed item set using feedback-based DCG.

\paragraph{Preference construction.}
Let $B$ contain $L$ exposed candidates from one request, and let $\Pi(B)$ denote their permutations. 
We use the same feedback score for WSFT weighting and permutation preference construction, setting \(\mathrm{Gain}(i)=w(i)\). 
These gains are treated as fixed item-level supervision labels across permutations.
We use the position-discounted gain~\cite{jarvelin2002cumulated} to define a feedback-based ranking criterion:
\begin{equation}
\operatorname{DCG}(Y)
=\sum_{t=1}^{L}
\frac{\operatorname{Gain}(y_t)}{\log_2(t+1)}
\label{eq:permutation_dcg}
\end{equation}

The strict preference-pair set is
\begin{equation}
\mathcal{P}(B)=\left\{
(Y^+,Y^-)\in\Pi(B)^2:
\operatorname{DCG}(Y^+) > \operatorname{DCG}(Y^-)
\right\}
\label{eq:permutation_pairs}
\end{equation}

With the candidate set fixed, the position discount gives greater weight to earlier positions, providing order-sensitive supervision that encourages items with higher observed gains to appear earlier. 
Permutations with equal DCG do not form preference pairs.
\paragraph{Preference optimization.}
For a permutation \(Y\), we compute its sequence log-probability using the next-item distribution defined in Eq.~(\ref{eq:rec logits}), without the tower adjustment \(\delta_{\phi, t}\):
\begin{equation}
    \label{eq:PermPO_s}
    \hat{s}_{\theta}(Y) = \sum_{t=1}^{L} \log p_{\theta}(y_t \mid y_{<t},x,C).
\end{equation}

The PermPO loss increases the sequence log-probability of the higher-DCG permutation relative to its paired lower-DCG permutation.
\begin{equation}
\mathcal{L}_{\mathrm{PermPO}}
=\frac{1}{|\mathcal{P}(B)|}
\sum_{(Y^+,Y^-)\in\mathcal{P}(B)}
-\log\sigma\left(
\hat{s}_{\theta}(Y^+)-\hat{s}_{\theta}(Y^-)
\right)
\label{eq:permpo_loss}
\end{equation}
where $\sigma$ is the sigmoid function.
Requests with no strict preference pairs contribute zero to this loss.
By computing sequence scores solely from the generator logits $z_{\theta,t}$, PermPO directly applies feedback-based relative ranking supervision to the generator used during serving, complementing the feedback-weighted next-item supervision provided by WSFT.

\subsubsection{Optimization}
\label{sec:optimization}
The overall training objective combines the NTP, WSFT, and PermPO losses:
\begin{equation}
\mathcal{L}
=\mathcal{L}_{\mathrm{NTP}}
+\lambda_{\mathrm{WSFT}}\mathcal{L}_{\mathrm{WSFT}}
+\lambda_{\mathrm{PermPO}}\mathcal{L}_{\mathrm{PermPO}},
\label{eq:joint_loss}
\end{equation}
where $\lambda_{\mathrm{WSFT}},\lambda_{\mathrm{PermPO}}\geq0$ control the contributions of WSFT and PermPO, respectively.

\begin{figure*}[t]
    \centering
    \includegraphics[width=\linewidth]{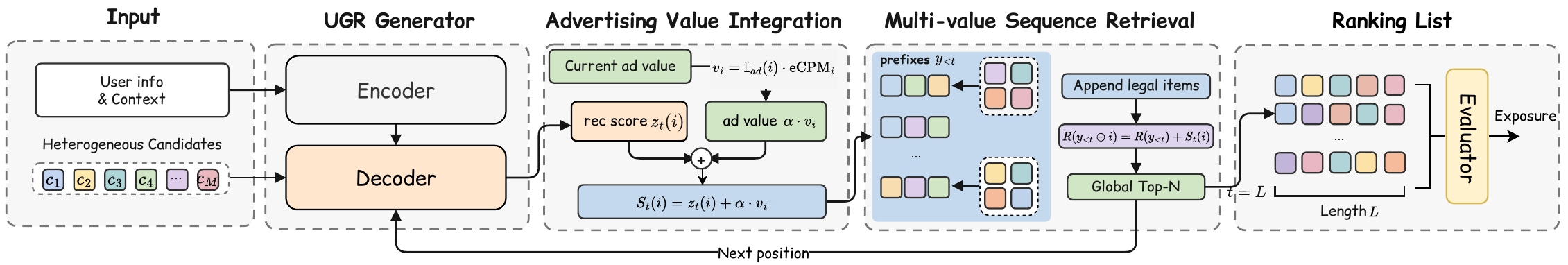}
    \caption{Online serving of UGR. 
    The generator scores heterogeneous candidates conditioned on user, context and the current prefix. 
    Recommendation scores are combined additively with current ad value.
    Multi-value sequence retrieval appends legal items and retains the top-$N$ prefixes by cumulative integrated score until length-$L$ ranking lists are generated.}
    \label{fig:infer}
\end{figure*}

\subsection{Online Serving}
\label{sec:online_serving}

During online serving (Fig.~\ref{fig:main}(c)), UGR removes the training-time advertising debiasing tower and integrates current ad value into the recommendation scores.
Multi-value sequence retrieval generates ranking lists, from which the existing downstream evaluator selects the final list for exposure.
The details are shown in Fig.~\ref{fig:infer}.

\subsubsection{Advertising Value Integration}
\label{sec:commercial_value}

For an ad \(i\) at decoding step \(t\), UGR directly combines the prefix-conditioned recommendation score with current ad value:
\begin{equation}
    S_t(i)=z_{\theta,t}(i)+\alpha\, \text{eCPM}_i
    \label{eq:serving_score}
\end{equation}
where $\mathrm{eCPM}_i$ is the current ad value of candidate $i$, and $\alpha>0$ controls its contribution. 
For ads, $\mathrm{eCPM}_i=q_i b_i$ under the value convention in Section~\ref{sec:ad_allocation_pricing}.
Under the fixed prefix \(y_{<t}\), the highest competing score for candidate $i$ is the highest score among the other legal candidates:
\begin{equation}
S_{-i,t}^{*}
=\max_{j\in C_t(y_{<t})\setminus\{i\}}S_t(j)
\label{eq:competing_score}
\end{equation}

With the prefix and competitive conditions fixed, the step-wise critical bid required for ad $i$ to win the current decoding step satisfies
\begin{align}
z_{\theta,t}(i)&+\alpha q_i b_{i,t}^{*}
=S_{-i,t}^{*}, \\
b_{i,t}^{*}
&=\frac{S_{-i,t}^{*}-z_{\theta,t}(i)}{\alpha q_i},
\alpha>0,\;q_i>0
\label{eq:slot_bid}
\end{align}

By separating historical ad exposure modeling during training from current ad value integration during serving, UGR incorporates advertising value into generative ranking while remaining compatible with the existing ad allocation and GSP pricing mechanisms.

\subsubsection{Multi-value Sequence Retrieval}
\label{sec:beam_search}

Multi-value sequence retrieval uses the integrated scores defined in Eq.~(\ref{eq:serving_score}) to search for ranking lists.
Encoder and candidate encodings are computed once and shared across search paths.
The generated prefix starts empty, while session tokens initialize the decoder context. 
At step $t$, each retained prefix is expanded with every candidate in $C_t(y_{<t})$, and the decoder scores each extension under its prefix.
Appending candidate $i$ to prefix $y_{<t}$ gives
\begin{equation}
R(y_{<t}\oplus i)=R(y_{<t})+S_t(i),
\qquad R(\varnothing)=0
\label{eq:beam_score}
\end{equation}
where $\oplus$ denotes appending a candidate to the prefix. 
Thus, each path is scored by the sum of its integrated scores across
decoding steps.

All extensions from the retained prefixes are ranked jointly, and the top $N$ legal paths are retained for the next step. 
For each retained path, the legal candidate set at the next step excludes candidates
already included in its prefix.
After \(L\) steps, multi-value sequence retrieval returns at most \(N\) ranking lists, from which the existing downstream list evaluator selects the final list for exposure.

\section{Experiments}
\label{sec:experiments}

\subsection{Experimental Setup}
\label{sec:experimental_setup}

\subsubsection{Datasets}
\label{sec:datasets}
We constructed two industrial datasets based on 30 consecutive days of recommendation logs from the Guess You Like recommendation service on the e-commerce platform Taobao's homepage.
The datasets represent two scenarios with different user intents. 
Table~\ref{tab:dataset_statistics} summarizes the dataset statistics.
Each sample corresponds to a recommendation request and includes user and context features, the full candidate set from the upstream stage, the logged exposure list, and observed post-exposure feedback.
The candidate set contains organic content, ads, and candidates from other businesses, without truncation or sampling. 
Exposure lists and feedback provide supervision for list generation and evaluation.
We split each dataset chronologically into training, validation, and test sets by request time.

\begin{table}[htbp]
\centering
\caption{Statistics of the two industrial datasets.}
\label{tab:dataset_statistics}
\small
\begin{tabular}{@{}cccc@{}}
\toprule
\textbf{Datasets} & \textbf{Users} & \textbf{Items} &  \textbf{Requests}  \\
\midrule
Scenario 1 & 54,682,294 & 118,068,621 & 1,938,791,731 \\
Scenario 2 & 24,151,394 & 98,072,780 & 237,500,048 \\
\bottomrule
\end{tabular}
\end{table}

\begin{table*}[h]
\centering
\caption{Offline performance on the two industrial datasets. Higher values are better. Bold and underlined values indicate the best and second-best reported results in each column, respectively.}
\label{tab:offline_comparison}
\setlength{\tabcolsep}{3pt}
\begin{tabular}{@{}l*{14}{r}@{}}
\toprule
\multirow{3}{*}[-1.4ex]{\textbf{Model}}
& \multicolumn{7}{c}{\textbf{Scenario 1}}
& \multicolumn{7}{c}{\textbf{Scenario 2}} \\
\cmidrule(lr){2-8}\cmidrule(lr){9-15}
& \multirow{2}{*}{\textbf{Hitrate}}
& \multicolumn{3}{c}{\textbf{IPV}}
& \multicolumn{3}{c}{\textbf{CVR}}
& \multirow{2}{*}{\textbf{Hitrate}}
& \multicolumn{3}{c}{\textbf{IPV}}
& \multicolumn{3}{c}{\textbf{CVR}} \\
\cmidrule(lr){3-5}\cmidrule(lr){6-8}
\cmidrule(lr){10-12}\cmidrule(lr){13-15}

& & \textbf{NDCG} & \textbf{MAP} & \textbf{AUC} & \textbf{NDCG} & \textbf{MAP} & \textbf{AUC}
& & \textbf{NDCG} & \textbf{MAP} & \textbf{AUC} & \textbf{NDCG} & \textbf{MAP} & \textbf{AUC} \\
\midrule
PRM \cite{prm} & 0.5295 & 0.3454 & 0.2906 & 0.5603 & 0.5305 & 0.4607 & 0.6018 & 0.4882 & 0.3998 & 0.3420 & 0.5626 & 0.5620 & 0.4886 & 0.6321 \\
NAR4Rec \cite{nar4rec} & 0.5050 & 0.3429 & 0.2827 & 0.5507 & 0.4361 & 0.3682 & 0.5680 & 0.4648 & 0.4002 & 0.3349 & 0.5760 & 0.4989 & 0.4342 & 0.6336 \\
GReF \cite{gref} & 0.5310 & 0.3459 & 0.2926 & 0.6036 & 0.5274 & 0.4744 & \underline{0.6894} & \underline{0.5220} & 0.4261 & 0.3642 & 0.6090 & 0.4856 & 0.4368 & 0.6882 \\
Dual-Rerank \cite{dualrerank} & \underline{0.5361} & 0.3451 & 0.2886 & 0.5731 & 0.5207 & 0.4632 & 0.6582 & 0.4998 & 0.4002 & 0.3393 & 0.5911 & 0.5617 & 0.5036 & 0.6743 \\
UniRank \cite{unirank} & \textbf{0.5668} & \underline{0.3496} & 0.2802 & 0.4799 & 0.4514 & 0.3650 & 0.4831 & \textbf{0.5343} & 0.4124 & 0.3230 & 0.4580 & 0.5096 & 0.4155 & 0.5269 \\
DEGR \cite{degr} & 0.5267 & 0.3474 & \underline{0.2947} & \underline{0.6055} & \underline{0.5314} & \underline{0.4773} & 0.6837 & 0.5099 & \underline{0.4264} & \underline{0.3684} & \underline{0.6338} & \underline{0.5807} & \underline{0.5227} & \underline{0.7014} \\
\textbf{UGR (Ours)} & 0.5295 & \textbf{0.3666} & \textbf{0.3169} & \textbf{0.6512} & \textbf{0.5394} & \textbf{0.4871} & \textbf{0.7023} & 0.4905 & \textbf{0.4325} & \textbf{0.3782} & \textbf{0.6714} & \textbf{0.5824} & \textbf{0.5264} & \textbf{0.7035} \\

\bottomrule
\end{tabular}
\end{table*}

\subsubsection{Baselines}
\label{sec:baselines}
We compare UGR with six representative ranking and generative ranking baselines, including PRM~\cite{prm}, NAR4Rec~\cite{nar4rec}, GReF~\cite{gref}, Dual-Rerank~\cite{dualrerank}, UniRank~\cite{unirank}, and DEGR~\cite{degr}.
Detailed descriptions are provided in Appendix~\ref{app:baselines}.

\subsubsection{Metrics}
\label{sec:metrics}

We evaluate each generated list using logged exposures and observed feedback for each request. 
Hitrate measures the overlap between generated lists and logged exposure lists.
NDCG, MAP, and AUC are reported separately for engagement (IPV) and transaction (CVR). 
For each feedback type, an item is labeled positive if the corresponding behavior was observed, and negative otherwise.
Higher values indicate better performance.
Detailed definitions of the metrics are provided in Appendix~\ref{app:Offline_Evaluation_Metrics}.

\subsubsection{Implementation Details}
\label{sec:implementation_details}
Production requests typically contain on the order of a thousand candidates across businesses.
Our implementation uses a configured total candidate capacity of 1,000 per request and generates four-item lists with a retrieval width of four.
It uses 256-dimensional token representations and eight attention heads in both the encoder and decoder.
The candidate encoder combines low-rank attention with a feed-forward sublayer, and the advertising debiasing tower is an MLP with one hidden layer of width 8. 
We train UGR with AdagradDecayV2 using a batch size of 128 and a learning rate of 0.01, with linear warmup from 0.001 over the first $10^6$ steps. 
The WSFT and PermPO loss weights are 1 and 15, respectively. 
Additional model and training configurations are provided in Appendix~\ref{app:implementation_details}.

\subsection{Offline Evaluation}
\label{sec:offline_comparison}

We compare UGR with the baseline methods on the two industrial datasets. 
Table~\ref{tab:offline_comparison} reports exposure agreement and ranking quality for engagement and transaction feedback.

In Scenario 1, UGR achieves the highest values on all six ranking metrics. 
It improves IPV NDCG by \textbf{4.86\%} over UniRank and IPV MAP by \textbf{7.53\%} over DEGR, the strongest baselines on the respective metrics. 
These improvements indicate better engagement-oriented ranking quality under both position-sensitive and precision-based evaluation. 
For transaction feedback, UGR improves CVR MAP by \textbf{2.05\%} over DEGR and also leads on the remaining CVR ranking metrics. 
Its gains therefore extend to both engagement and transaction objectives rather than being concentrated on a single feedback target.
UGR also achieves the best performance across all six ranking metrics in Scenario 2, showing consistent improvements across the two industrial scenarios.
UniRank achieves the highest Hitrate in both scenarios, whereas UGR leads on feedback-based ranking quality. 
This distinction indicates that stronger agreement with logged exposures does not necessarily translate into better ranking according to observed feedback. 
Together, these results demonstrate UGR’s effectiveness in ranking heterogeneous candidates for both engagement and transaction value.

\subsection{Online A/B Test}
\label{sec:online_ab_test}
We conduct an online A/B test on Taobao's Guess You Like recommendation service, comparing UGR with the production baseline.
The UGR treatment group covers at least 5\% of the recommendation service traffic, and the experimental deployment lasts for more than 14 days.
The downstream evaluator and the remaining serving pipeline are kept unchanged between the control and treatment groups. 

\begin{table}[t]
\centering
\caption{
Relative improvements over the production baseline. All results are statistically significant with $p<0.05$.
}
\label{tab:online_results}
\setlength{\tabcolsep}{3pt}
\begin{tabular}{@{}ccccc@{}}
\toprule
\textbf{Core IPV} & \textbf{Click UV} & \textbf{UCTR} & \textbf{Transaction Volume} & \textbf{GMV}\\
\midrule
+1.76\% & +1.32\% & +0.45\% & +3.29\% & +2.10\%  \\
\bottomrule
\end{tabular}
\end{table}

We evaluate user engagement using Core IPV, Click UV, and UCTR, and transaction outcomes using Transaction Volume and GMV. 
As shown in Table~\ref{tab:online_results}, UGR achieves statistically significant improvements across all five metrics. 
In particular, the gains in Core IPV, Click UV, and UCTR demonstrate improved user engagement, while the improvements in Transaction Volume and GMV show that these gains extend to transaction outcomes. 
The Ad Cost changes by +0.08\% while the Ad PVR changes by -0.93\%, with both metrics satisfying the ad guardrail requirements.
The average inference latency of UGR serving increases by only 12 ms, meeting the low-latency requirements of large-scale real-time systems.
Detailed metric definitions are provided in Appendix~\ref{app:online_metrics}.

\subsection{Analysis of Unified Ranking}
\label{sec:generation-space}
We examine organic-candidate truncation offline and compare unified generation with single-business generation followed by cross-business blending online, with the output length fixed.

\subsubsection{Organic Candidate Truncation}
\begin{figure}[t]
    \centering
    \includegraphics[width=\linewidth]{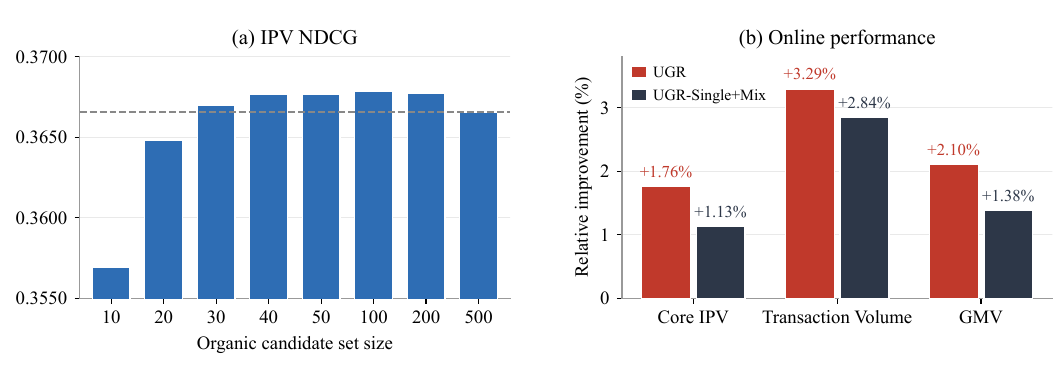}
    \caption{Analysis of unified ranking.
(a) Effect of organic-candidate truncation on offline IPV NDCG, with candidates from other businesses held fixed. The dashed line marks the result with the full organic candidate set ($K=500$).
(b) Online improvements of UGR and UGR-Single+Mix over the same production baseline.}
    \label{fig:generation-space}
\end{figure}

We vary only the organic portion of the heterogeneous candidate set by retaining the upstream top-$K$ organic candidates, with $K=500$ covering the full organic set in this experiment.
Candidates from other businesses remain unchanged. 
For each $K$, we train a separate model and evaluate it on the same test requests.
Fig.~\ref{fig:generation-space}(a) shows that restricting the set to 10 or 20 organic candidates reduces IPV NDCG, whereas the full-organic-set result remains within approximately 0.5\% of the results obtained with $K$ between 30 and 200. 
Thus, retaining the full organic candidate set achieves comparable offline IPV NDCG to the tested configurations with \(K\) between 30 and 200.

\subsubsection{Comparison with Staged Ranking}
We compare UGR with UGR-Single+Mix, which restricts UGR to organic-only lists and leaves cross-business blending to the existing mixrank, whereas UGR generates heterogeneous lists directly.
Both models are deployed concurrently with the production baseline, each serving at least 5\% of traffic for more than 14 days. 
The serving configuration, the downstream evaluator, and all ad constraints are identical across the three groups.

Fig.~\ref{fig:generation-space}(b) reports the improvements over the same production baseline. 
UGR-Single+Mix improves Core IPV, Transaction Volume, and GMV by 1.13\%, 2.84\%, and 1.38\%, while UGR improves them by 1.76\%, 3.29\%, and 2.10\%. 
UGR also significantly outperforms UGR-Single+Mix in the direct comparison.
The staged variant already benefits from UGR's single-business ranking capability, but the subsequent cross-business blending step changes item positions and the surrounding list context.
These results support determining cross-business blending and item ranking within a single generation process.

\subsection{Ablation Study}
\label{sec:ablation_study}

\subsubsection{Training Components}
We examine the contributions of UGR's training components on Scenario 1. 
Table~\ref{tab:ablation} compares full UGR with NTP-only and variants without WSFT, PermPO, or the advertising debiasing tower.

\begin{table}[htbp]
\centering
\caption{Ablation results for UGR on Scenario 1.}
\label{tab:ablation}
\small
\setlength{\tabcolsep}{3pt}
\begin{tabular}{@{}lccccccc@{}}
\toprule
\multirow{2}{*}[-0.4ex]{\textbf{Model}}
& \multirow{2}{*}[-0.4ex]{\textbf{Hitrate}}
& \multicolumn{3}{c}{\textbf{IPV}}
& \multicolumn{3}{c}{\textbf{CVR}} \\
\cmidrule(lr){3-5}\cmidrule(lr){6-8}
& & \textbf{NDCG} & \textbf{MAP} & \textbf{AUC} & \textbf{NDCG} & \textbf{MAP} & \textbf{AUC} \\
\midrule
UGR (full) & 0.5295 & 0.3666 & 0.3169 & 0.6512 & 0.5394 & 0.4871 & 0.7023 \\
NTP-only & 0.5291 & 0.3382 & 0.2819 & 0.5706 & 0.5122 & 0.4566 & 0.6630 \\
w/o WSFT & 0.5255 & 0.3539 & 0.3059 & 0.6520 & 0.5375 & 0.4864 & 0.7069 \\
w/o PermPO & 0.5296 & 0.3453 & 0.2883 & 0.5738 & 0.5137 & 0.4576 & 0.6614 \\
w/o tower & 0.5227 & 0.3614 & 0.3121 & 0.6491 & 0.5256 & 0.4716 & 0.6924 \\
\bottomrule
\end{tabular}
\end{table}

UGR improves all six ranking metrics over NTP-only training, with comparable Hitrate.
WSFT improves NDCG and MAP, particularly for IPV, with a slight reduction in AUC.
Removing PermPO lowers all six ranking metrics, indicating its contribution to learning relative ranking preferences.
Together, these results demonstrate the complementary roles of feedback-weighted supervision and permutation preference optimization.

\subsubsection{Advertising Debiasing Tower}
\label{sec:Advertising Debiasing Tower}
\begin{figure}[t]
    \centering
    \includegraphics[width=\linewidth]{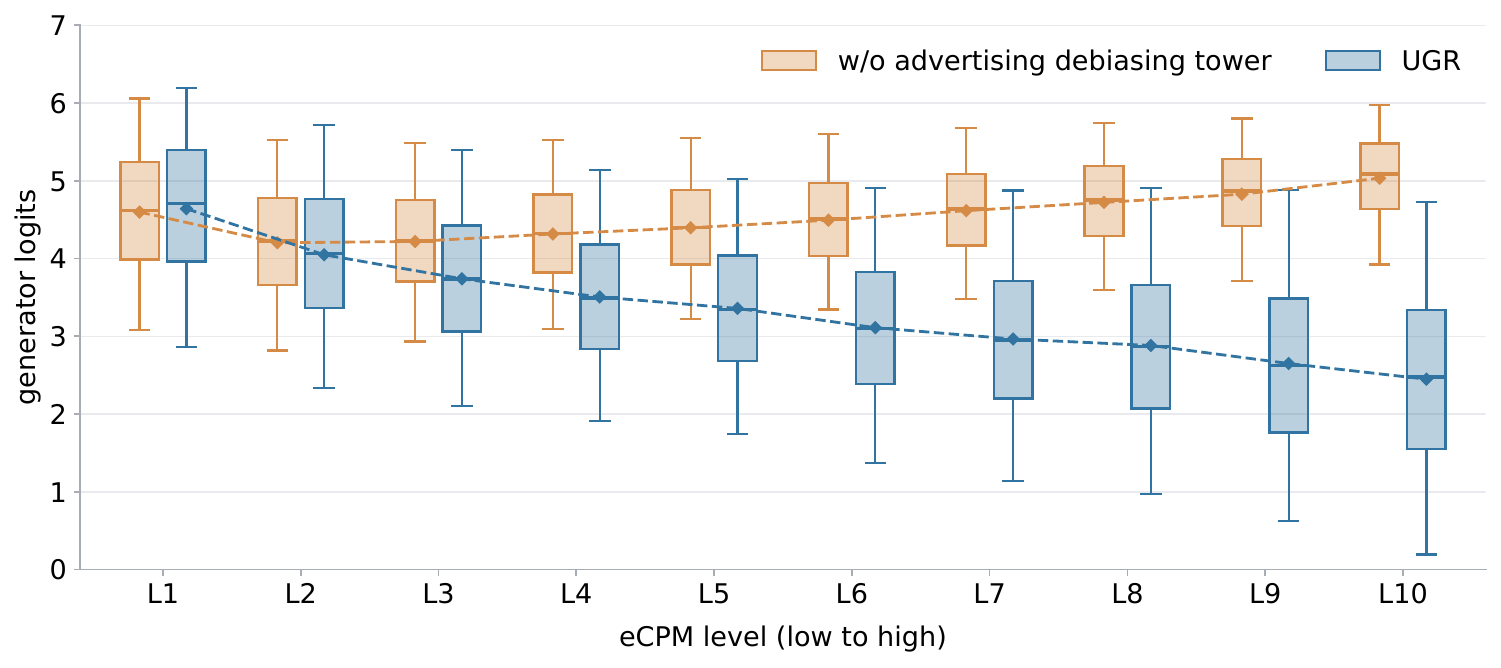}
    \caption{Generator logit distributions of UGR and UGR without the advertising debiasing tower across ad eCPM intervals. Boxes show the interquartile range, center lines indicate medians, whiskers mark the 5th and 95th percentiles, and dashed lines connect mean values.}
    \label{fig:debiasing_analysis}
\end{figure}

Removing the debiasing tower reduces Hitrate and ranking quality.
To further examine this effect, we analyze how the generator logits from UGR and UGR without the debiasing tower vary with ad eCPM, as shown in Fig.~\ref{fig:debiasing_analysis}.

Without the tower, the generator logits $z_t$ exhibit an overall positive correlation with eCPM ($r=0.195$).
For UGR, the correlation between generator logits $z_t$ and eCPM becomes negative ($r=-0.313$), whereas the tower adjustment $\delta_t$ is positively correlated with eCPM ($r=0.727$, see Appendix~\ref{app:Advertising Debiasing Tower} for the component distributions and their sum).
These contrasting associations suggest that the tower separates eCPM-related effects from the generator logits.
During training, candidate probabilities depend on the sum of the generator logit and  tower adjustment, allowing a larger adjustment to offset a lower generator logit.
Together with the ablation results, these patterns support the effectiveness of modeling ad exposure effects through a separate training-time tower.

\subsection{Hyperparameter Analysis}
\label{sec:hyperparameter_analysis}

We analyze the effects of the WSFT and PermPO loss weights by varying $\lambda_{\mathrm{WSFT}}$ and $\lambda_{\mathrm{PermPO}}$ independently.
Fig.~\ref{fig:hyperparameter-main} reports Hitrate and NDCG.
Complete results are provided in Appendix~\ref{app:Additional_Hyperparameter_Analysis}.

\begin{figure}[htbp]
    \centering
    \includegraphics[width=\linewidth]{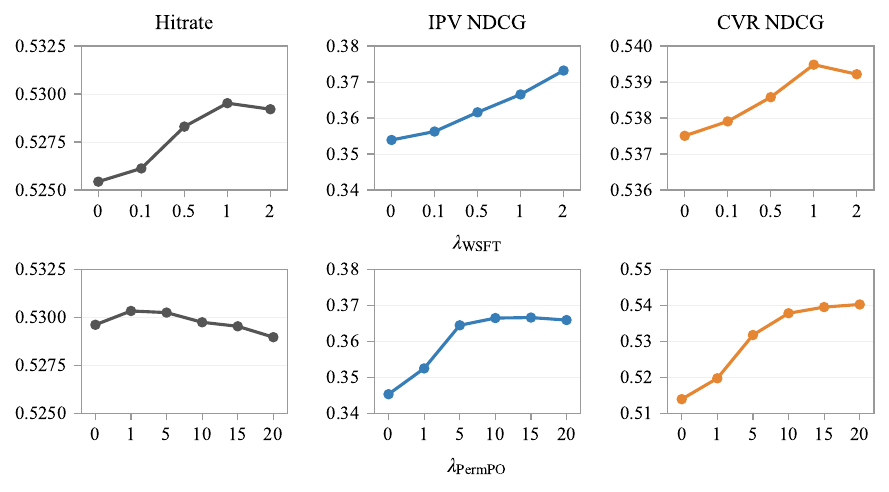}
    \caption{Effects of the WSFT and PermPO loss weights. }
    \label{fig:hyperparameter-main}
\end{figure}

\subsubsection{WSFT Weight}
\label{sec:wsft_weight}
Increasing $\lambda_{\mathrm{WSFT}}$ consistently improves IPV NDCG, whereas CVR NDCG reaches its maximum at
$\lambda_{\mathrm{WSFT}}=1$. 
Hitrate follows a similar pattern, increasing up to a weight of 1 and decreasing slightly thereafter.
We set $\lambda_{\mathrm{WSFT}}=1$, which achieves the highest Hitrate and CVR NDCG while retaining most of the improvement in IPV NDCG.

\subsubsection{PermPO Weight}
\label{sec:permpo_weight}
Increasing $\lambda_{\mathrm{PermPO}}$ substantially improves both IPV and CVR NDCG. 
Most of the IPV gain is obtained at a weight of 15, after which the performance remains nearly stable, while CVR NDCG continues to improve more gradually.
In contrast, Hitrate peaks at a small weight and decreases as the PermPO weight increases. 
We set $\lambda_{\mathrm{PermPO}}=15$, which captures the ranking gains before they largely saturate.

\subsection{Scaling Analysis}
\label{sec:Scaling_analysis}

We vary the token dimension over $\{64, 128, 256, 512, 1024\}$ to study how representation capacity affects UGR. 
Fig.~\ref{fig:scaling-main} reports NDCG for IPV and CVR. 
Complete results are provided in Appendix~\ref{sec:additional-scaling-results}.

\begin{figure}[htbp]
    \centering
    \includegraphics[width=\linewidth]{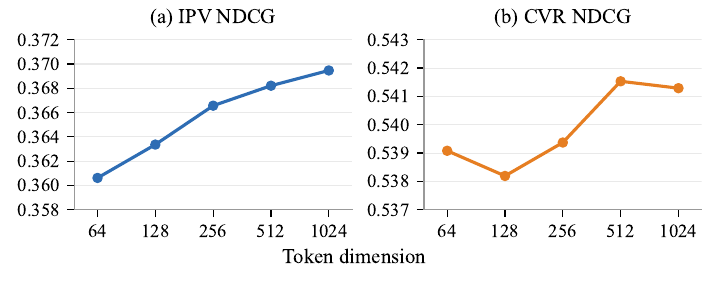}
    \caption{
        Effect of token dimension on IPV and CVR NDCG.
    }
    \label{fig:scaling-main}
\end{figure}

IPV NDCG increases with token dimension, whereas CVR NDCG peaks at 512 and decreases slightly at 1024. 
Overall, these results suggest that increasing representation capacity can improve UGR's ranking performance.

\section{Conclusions}
\label{sec:conclusions}
We presented UGR, a framework that directly generates ranking lists over the full request-level heterogeneous, multi-business candidate set received from the upstream stage.
UGR extends exposure-sequence imitation with complementary feedback supervision at the item and list levels.
By separating historical ad exposure modeling during training from current ad value integration during serving, UGR enables generative ranking to operate within the existing ad allocation and GSP pricing rules.
Experiments on two industrial datasets and large-scale online A/B tests consistently validate the effectiveness of UGR. 
The results demonstrate the benefits of unified generative ranking for improving recommendation quality and business outcomes, as well as its practicality for industrial deployment.

\bibliographystyle{ACM-Reference-Format}
\bibliography{ref}

\appendix

\section{Implementation Details}
\label{app:implementation_details}

This section details the model architecture, training configuration, and serving procedure of UGR.

\subsection{Model Architecture}
\label{app:model_architecture}

Table~\ref{tab:implementation_configuration} summarizes the model, training, and serving settings.

\begin{itemize}
    \item \textbf{Request context encoder.}
     Each encoder layer consists of masked self-attention over the feature and behavior tokens, followed by a position-wise feed-forward network, with a residual connection and RMSNorm around each sublayer. A learned token-axis projection then compresses the encoded sequence into a single request context token.

    \item \textbf{Candidate encoding.}
    Candidate features are projected by an MLP with hidden widths of 512 and 256. Candidate interactions are modeled by one low-rank attention block with eight latent key--value components, followed by a two-layer feed-forward network
    with widths of 256 and 256.

    \item \textbf{Autoregressive decoder.}
     The decoder block applies causal self-attention over the session and selected-candidate tokens, cross-attention to the request context token, and a two-layer feed-forward network. Each sublayer is wrapped with a residual connection and RMSNorm.

    \item \textbf{Advertising debiasing tower.}
    The advertising debiasing tower has a hidden width of eight and is used only during training.
    The output layer and position offsets are initialized to zero, so the tower initially leaves the generator logits unchanged.

    \item \textbf{Activation and normalization.}
    We use LeakyReLU activation and RMSNorm with $\epsilon=10^{-8}$. 
\end{itemize}

\subsection{Training Configuration}
\label{app:training_configuration}

\begin{itemize}
    \item \textbf{Optimization.}
    We use AdagradDecayV2 with a training batch size of 128.
    The learning rate increases linearly from 0.001 to 0.01 over the first $10^6$ global steps and remains at 0.01 thereafter. 

    \item \textbf{Training objective.}
    NTP, WSFT, and PermPO are optimized jointly, with coefficients 1, $\lambda_{\mathrm{WSFT}}=1$ and $\lambda_{\mathrm{PermPO}}=15$, respectively.

    \item \textbf{Feedback definition and weighting.} We use two binary feedback signals. $F_{\text{IPV}}=1$ if item $i$ receives a click or has a browsing duration exceeding 3 seconds, and $F_{\text{CVR}}=1$ if a transaction is observed for the item; otherwise, the corresponding signal is zero.
    WSFT and PermPO share the same feedback score: $w(i)=Gain(i)=F_{\text{IPV}}+2\,F_{\text{CVR}}$. This corresponds to $\eta=1,\beta_{\text{IPV}}=1$, and $\beta_{\text{CVR}}=2$ in Eq.~(\ref{eq:weight feedback}). 
    The two contributions are summed directly, without clipping or normalization.

\end{itemize}

\begin{table}[t]
    \centering
    \caption{Model architecture, training and serving settings.}
    \label{tab:implementation_configuration}
    \small
    \setlength{\tabcolsep}{4pt}
    \begin{tabular}{@{}p{0.64\columnwidth}p{0.28\columnwidth}@{}}
        \toprule
        Setting & Value \\
        \midrule
        \multicolumn{2}{@{}l}{\textbf{Model architecture}} \\
        Token representation dimension & 256 \\
        Request encoder attention layers & 1 \\
        Candidate embedding MLP widths & $512\rightarrow256$ \\
        Candidate-attention blocks & 1 \\
        Candidate-attention rank & 8 \\
        Candidate FFN widths & $256\rightarrow256$ \\
        Decoder FFN widths & $256\rightarrow256$ \\
        Activation & LeakyReLU \\
        Encoder / decoder normalization & RMSNorm \\
        \midrule
        \multicolumn{2}{@{}l}{\textbf{Training configuration}} \\
        Optimizer & AdagradDecayV2 \\
        Training batch size & 128 \\
        Learning rate after warmup & 0.01 \\
        Warmup starting learning rate & 0.001 \\
        Warmup steps & $10^6$ \\
        $\lambda_{\mathrm{WSFT}}$ & 1 \\
        $\lambda_{\mathrm{PermPO}}$ & 15 \\
        \midrule
        \multicolumn{2}{@{}l}{\textbf{Serving configuration}} \\
        Total candidate capacity $M$ & 1,000 \\
        Number of returned candidate lists $N$ & 4 \\
        Items per generated list $L$ & 4 \\
        \bottomrule
    \end{tabular}
\end{table}

\subsection{Serving and Deployment}
\label{app:serving_deployment}

\begin{itemize}
    \item \textbf{Multi-value sequence retrieval.}
    Candidate extensions are scored using the online value integration described in Section~\ref{sec:online_serving}.
    Partial sequences are ranked by their cumulative integrated scores, and the four highest-scoring extensions across all retained prefixes are kept at each step.
    Padded candidates and items already selected in the current prefix are masked. 
    After four decoding steps, UGR returns four ranking lists, each containing four distinct items.

    \item \textbf{Computation reuse.}
    The request encoding, candidate embeddings, and candidate-attention outputs are computed once per request and reused throughout retrieval.
    The decoder recomputes the current prefix at each step.

    \item \textbf{Production integration.}
    UGR is deployed downstream of the candidate retrieval and scoring modules.
    Its generated lists are ranked by the same downstream evaluator used by the production baseline, which selects the final list for exposure.
\end{itemize}

\subsection{Baseline Details}
\label{app:baselines}
We briefly describe the six baselines used in our experiments.
\begin{itemize}
    \item 	\textbf{PRM}~\cite{prm} uses personalized self-attention to model candidate interactions for reranking.
    \item 	\textbf{NAR4Rec}~\cite{nar4rec} generates lists in parallel using unlikelihood training and contrastive decoding.
    \item 	\textbf{GReF}~\cite{gref} combines autoregressive generation, Rerank-DPO, and ordered multi-token prediction.
    \item 	\textbf{Dual-Rerank}~\cite{dualrerank} combines non-autoregressive generation with sequence distillation and list-wise utility optimization.
    \item 	\textbf{UniRank}~\cite{unirank} unifies autoregressive and non-autoregressive ranking through confidence-driven iterative denoising.
    \item 	\textbf{DEGR}~\cite{degr} balances immediate and exploration value through dual exploration and reward-weighted preference optimization.
\end{itemize}

All methods use the same chronological data splits, request-time features, full upstream candidate sets, and output list length.
We retain each baseline's architecture, training objective, and decoding mechanism, adapting only its input and output interfaces to the heterogeneous candidate-ranking task.
Hyperparameters are tuned independently on the validation set, and model selection does not use test results. 
All methods are evaluated using the same feedback-label construction and metric implementations.

\section{Evaluation Metrics}
\label{app:evaluation-metrics}

This section defines the metrics used in our offline experiments and online A/B tests.

\subsection{Offline Evaluation Metrics}
\label{app:Offline_Evaluation_Metrics}
Following previous works~\cite{dualrerank, unirank, degr, gref, degre}, the offline metrics are computed as follows: 

\begin{itemize}
  \item \textbf{Hitrate} measures the exposure agreement between the generated ranking list and the logged exposure list.
  \item \textbf{NDCG} evaluates position-aware ranking quality, assigning greater importance to positive-feedback items ranked near the top.
  \item \textbf{MAP} evaluates the overall ranking quality of positive-feedback items in the generated list.
  \item \textbf{AUC} evaluates the relative ranking quality between positive- and negative-feedback items.
\end{itemize}

NDCG, MAP, and AUC are reported separately for IPV and CVR.
For IPV, an item is assigned a label of 1 if it receives a click or has a browsing duration exceeding 3 seconds. 
For CVR, an item is assigned a label of 1 if a transaction is observed. 
Higher values of all four metrics indicate better performance.

\subsection{Online A/B Test Metrics}
\label{app:online_metrics}

We define the online metrics reported in our A/B tests as follows.

\begin{itemize}
    \item \textbf{Core IPV (Item Page Views)} counts item-detail-page visits attributed to the evaluated recommendation service, restricted to the platform-defined core item scope.

    \item \textbf{Click UV} measures the approximate number of distinct users with at least one valid click on a recommended item.

    \item \textbf{UCTR (User Click-Through Rate)} is the ratio of Click UV to Exposure UV, where Exposure UV counts distinct users with at least one valid recommendation exposure.

    \item \textbf{Transaction Volume} counts completed orders attributed to the evaluated recommendation service.

    \item \textbf{GMV (Gross Merchandise Volume)} measures the total value of paid orders attributed to the evaluated recommendation service.

    \item \textbf{Ad PVR } measures the proportion of ad items among all displayed items. It gauges the extent of ad exposure and serves as a guardrail for monitoring changes in ad impressions.
    
    \item \textbf{Ad Cost} measures advertiser spend across the overall recommendation service and serves as a guardrail for advertising expenditure.

\end{itemize}

For each metric $M$, we report its relative improvement over the production baseline as $\Delta M = (M_{\mathrm{UGR}} / M_{\mathrm{baseline}} - 1) \times 100\%$.

\section{Additional Results}
\label{app：Additional_Results}
This section presents the complete results of the advertising debiasing tower analysis in Section~\ref{sec:Advertising Debiasing Tower}, hyperparameter analysis in Section~\ref{sec:hyperparameter_analysis}, and token-dimension scaling analysis in Section~\ref{sec:Scaling_analysis}.

\subsection{Advertising Debiasing Tower}
\label{app:Advertising Debiasing Tower}

\begin{figure}[htbp]
    \centering
    \includegraphics[width=\linewidth]{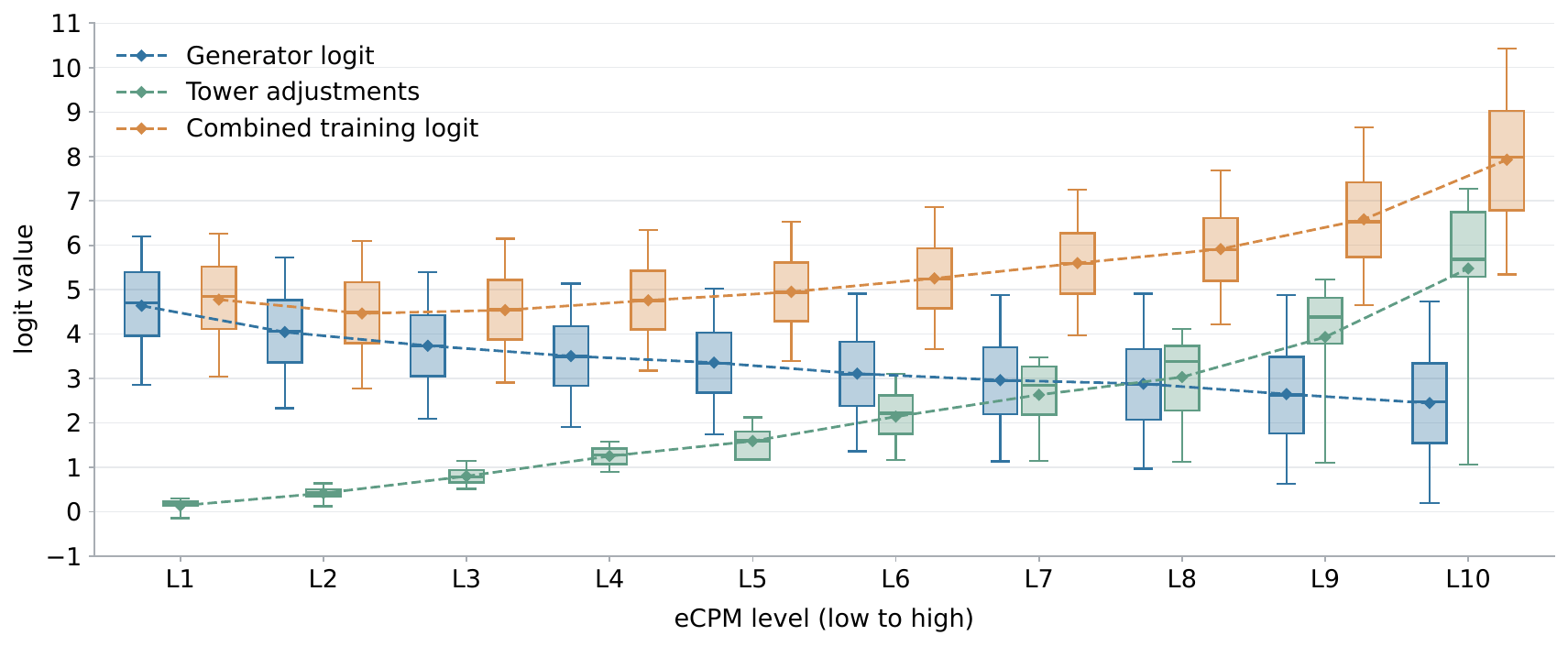}
    \caption{Distributions of UGR generator logits, tower adjustments, and combined training logits across eCPM intervals.
    Boxes show the interquartile range, center lines indicate medians, whiskers mark the 5th and 95th percentiles, and dashed lines connect mean values.}
    \label{fig:adt_full}
\end{figure}

Fig.~\ref{fig:adt_full} further presents the distributions of the UGR generator logits \(z_t(i)\), the tower adjustments \(\delta_t(i)\) produced by the debiasing tower, and combined training logits \(z_t(i)+\delta_t(i)\) across ten eCPM intervals ordered from low to high. 
Pearson correlations are computed between eCPM and each of these three quantities.

\begin{figure*}[t]
    \centering
    \includegraphics[width=0.9\linewidth]{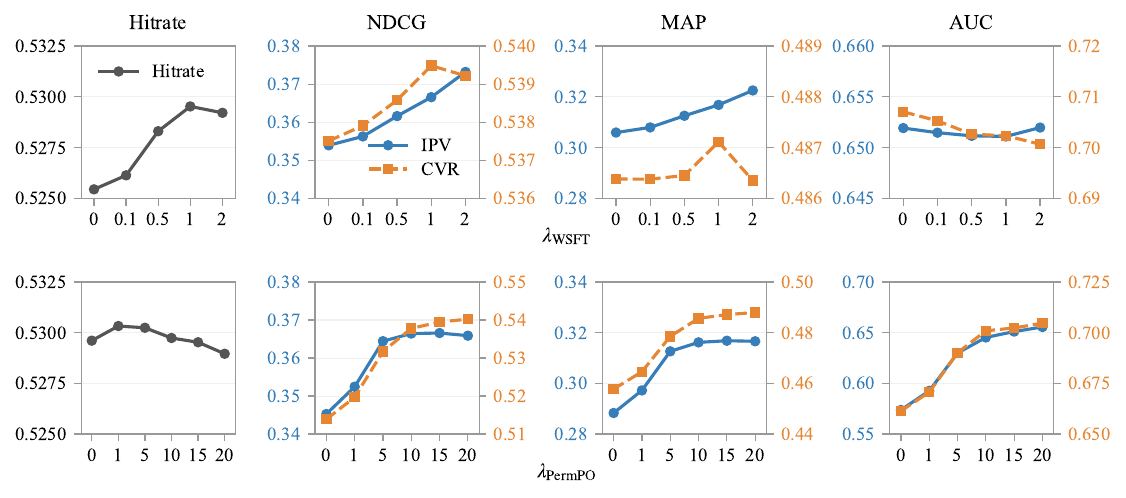}
    \caption{Complete hyperparameter analysis of the WSFT and PermPO loss weights.
    The top and bottom rows vary \(\lambda_{\mathrm{WSFT}}\) and \(\lambda_{\mathrm{PermPO}}\), respectively. 
    The columns report Hitrate, NDCG, MAP, and AUC.
    Hitrate is shown in gray; in the remaining columns, the blue and orange curves denote IPV and CVR performance, respectively.
    }
    \label{fig:hyperparameter-full}
\end{figure*}
\begin{figure*}[t]
    \centering
    \includegraphics[width=0.9\textwidth]{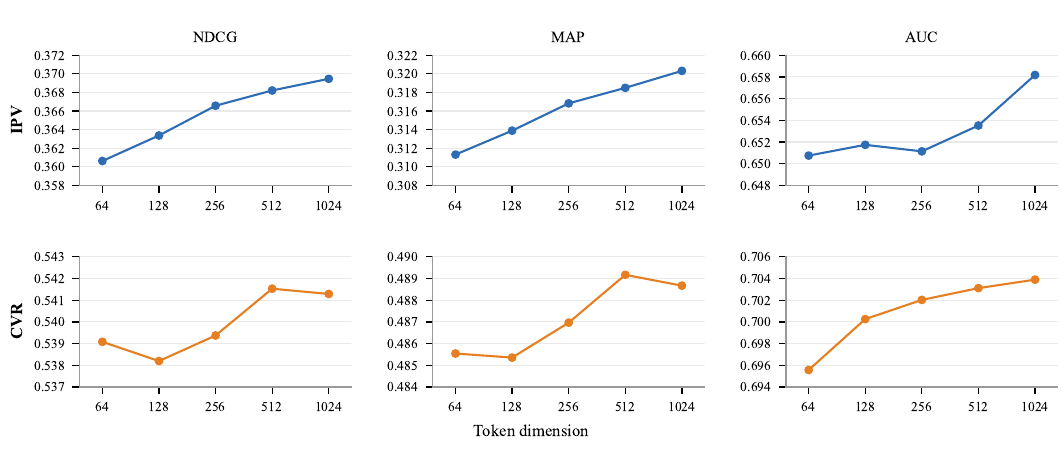}
    \caption{
        Complete scaling results across token dimensions.
        The top and bottom rows report performance for IPV and CVR, respectively. 
        From left to right, the columns show NDCG, MAP, and AUC.
    }
    \label{fig:scaling-full}
\end{figure*}

As the eCPM level increases, the generator logits decrease, while the tower adjustments increase. 
The combined logits show a slight initial decline followed by an upward trend at higher eCPM levels, where the increase in tower adjustments outweighs the decrease in generator logits. 
These patterns indicate that eCPM-related variation is reflected primarily in the tower adjustments rather than in the generator logits. 
During serving, only the generator logits are retained.

\subsection{Hyperparameter Analysis}
\label{app:Additional_Hyperparameter_Analysis}

Fig.~\ref{fig:hyperparameter-full} reports the complete hyperparameter results, including MAP and AUC. 

MAP exhibits trends similar to NDCG for both IPV and CVR. 
As \(\lambda_{\mathrm{WSFT}}\) increases, IPV AUC remains relatively stable while CVR AUC decreases slightly. 
Increasing $\lambda_{\mathrm{PermPO}}$ generally improves AUC, with the gains becoming smaller at larger weights.

\subsection{Scaling Performance}
\label{sec:additional-scaling-results}

Fig.~\ref{fig:scaling-full} reports the complete scaling results for NDCG, MAP, and AUC for both IPV and CVR.

For IPV, MAP improves consistently with the token dimension, while AUC shows an overall upward trend with minor intermediate variation. 
For CVR, MAP follows a pattern similar to NDCG, reaching its highest value at a dimension of 512 and decreasing slightly at 1024. 
In contrast, CVR AUC improves across the evaluated dimensions.

Overall, the effect of increasing representation capacity depends on both the ranking target and the evaluation metric.
Larger token dimensions generally improve IPV performance, while the improvements in CVR ranking quality become less pronounced at larger dimensions.
We use a token dimension of 256 in the main experiments to balance ranking quality and serving cost.

\end{document}